\documentclass[prb,showpacs,superscriptaddress,twocolumn,notitlepage]{revtex4-1}

\usepackage{graphicx, subfigure}

\usepackage{amsmath}
\usepackage{amsfonts}
\usepackage{graphicx,subfigure}

\usepackage{color}
\usepackage[usenames,dvipsnames]{xcolor}

\usepackage{natbib}
\usepackage[final=true]{hyperref}

\begin{document}
\title{Substrate-induced reduction of graphene thermal conductivity}

\author{S.~V.~Koniakhin}
\email{kon@mail.ioffe.ru}
\affiliation{St. Petersburg Academic University, Khlopina 8/3, 194021 St. Petersburg, Russia}
\affiliation{Ioffe Physical-Technical Institute of the Russian Academy of Sciences, 194021 St.~Petersburg, Russia}

\author{O.~I.~Utesov}
\email{utiosov@gmail.com}
\affiliation{Petersburg Nuclear Physics Institute NRC "Kurchatov Institute", Gatchina, St.\ Petersburg 188300, Russia}

\author{I.~N.~Terterov}
\affiliation{St. Petersburg Academic University, Khlopina 8/3, 194021 St. Petersburg, Russia}

\author{A.~V.~Nalitov}
\affiliation{School of Physics and Astronomy, University of Southampton, Southampton SO17 1BJ, United Kingdom}

\begin{abstract}

We develop the theory of heat conductivity in supported graphene, accounting for coherent phonon scattering on disorder induced by an amorphous substrate.
We derive spectra for in-plane and out-of-plane phonons in the framework of Green's function approach. The energetic parameters of the theory are obtained using the molecular dynamics simulations for graphene on SiO$_2$ substrate. The heat conductivity is calculated by the Boltzmann transport equation.
We find that the interaction with the substrate drastically reduces the phonon lifetime and completely suppresses the contribution of ZA phonons to the heat conductivity. As a result, the total heat conductivity is reduced by several times, which matches with the tendency observed for available experimental data.
The considered effect is important for managing thermal properties of the graphene-based electronic devices.
\end{abstract}

\maketitle

\section{Introduction}


Due to its outstanding properties, graphene, a honeycomb monolayer of carbon atoms, attracts great attention as a material for application in the future nanoscale electronics. The mechanical \cite{frank2007mechanical,lee2008measurement}, electronic \cite{RevModPhys.81.109,RevModPhys.83.407,PhysRevLett.98.186806} properties of graphene, various radiation driven effects \cite{PhysRevB.86.115301,Glazov2014101,PhysRevLett.107.276601} and thermoelectric \cite{dollfus2015thermoelectric} phenomena in graphene are among the hottest topics of recent studies in condensed matter physics.

Along with other carbon-based materials, graphene reveals extremely high thermal conductivity $
\kappa$. The room temperature thermal conductivity of suspended graphene reaches 5000 W\,m$^{-1}$K$^{-1}$ \cite{balandin2011thermal,nika2012two,1509.07762,0953-8984-28-48-483001}.
For application of graphene as a component of electronic devices and controlling the carrier density it is commonly gated. This requires close contact between graphene sheet and dielectric substrate like SiO$_2$.

Electron scattering on the surface charged impurities \cite{PhysRevLett.98.186806,doi:10.1143/JPSJ.75.074716}, surface corrugations \cite{doi:10.1021/nl070613a,PhysRevLett.100.016602,Katsnelson195}, atomic steps \cite{AtomicSteps} and surface polar phonons \cite{Chen2008,PhysRevB.77.195415,PhysRevB.81.195442} reduces electric conductivity of graphene. The same effect is expected for heat conductivity. Accounting for graphene layers and conductive traces as heat sinks requires a quantitative estimate of this effect. Also the estimations of acoustic phonon lifetime and mean free path are obligate for calculating phonon drag contributions to thermopower in graphene\cite{PhysRevB.82.155410,koniakhin2013phonon}.

Previously the effect of substrate on thermal properties of graphene was investigated experimentally, from theoretical standpoint and by molecular dynamics simulations. The experimental studies of heat conductivity in suspended graphene are based on the Raman optothermal method \cite{cai2010thermal,chen2010raman,ghosh2008extremely,balandin2008superior,faugeras2010thermal,chen2012thermal}, and reveal values above 2000\, Wm$^{-1}$K$^{-1}$. Investigation of the supported graphene thermal properties is based on the electrical heating and give several times lower values of thermal conductivity \cite{seol2010two}. Other works, where the supported graphene is studied, focus mainly on graphene electrical properties with heat conductivity being a secondary result \cite{murali2009breakdown,liao2011thermally}. The table in Ref. \cite{nika2012two} summarizes current studies on graphene thermal conductivity.

The heat conductivity is governed by the phonon dispersion and relaxation process \cite{balandin2011thermal,nika2012two}. The theoretically investigated mechanisms of phonon relaxation include boundary scattering, point defect scattering and anharmonic processes \cite{6235226,nika2009phonon,nika2009lattice,klemens1994thermal}. The effect of the substrate is out of consideration in most current theoretical works. In the supplementary online materials of Ref. \cite{seol2010two}, the spot contact model within Fermi golden rule formalism developing the Klemens approach \cite{klemens1955scattering} for phonon scattering, was employed for accounting for a substrate. However, the case of randomly distributed defects with close to atomic concentration needs a theory based on Green's function to account for coherent scattering.

The molecular dynamics (MD) is also used for investigating thermal conductivity of graphene \cite{zhang2011thermal,wei2011plane,zhu2016phonons,PhysRevB.90.195209} and carbon nanotubes \cite{PhysRevB.84.165418,PhysRevLett.99.255502}. Such studies include modeling of phonon of graphene on amorphous substrate. Ref.\cite{MDgraphene} is devoted to molecular dynamics simulations of graphene on SiO2 substrate and show that van der Waals interaction reduces relaxation time of intrinsic acoustic phonons in graphene. The non-equilibrium MD simulation demonstrates reduction of thermal transport in graphene on SiO$_{2}$ \cite{ong2011effect}. Although MD simulation is a promising and powerful tool for investigating vibrational properties of solids, it is necessary to compare the obtained results with independent analytical approach.

Here we address the effect of amorphous substrate on the phonon dispersion and lifetimes in graphene within the Green's function formalism\cite{gobel1998effects}. The parameters of the perturbing substrate-induced van der Waals potential required by the developed analytical theory were obtained with MD simulations. In-plane (LA,TA) and flexural (ZA) phonons were considered and corresponding contributions to heat conductivity were obtained within theBoltzmann transport equation (BTE) approach. The optical phonons were not considered due to their small occupation number at actual temperatures, which results in low contribution to thermal conductivity (see Refs. \cite{PhysRevB.82.115427,spectralHeatCond}).

The rest of the paper is organized as follows. In the beginning of Section \ref{Theorygeneral} we introduce the general form of graphene Hamiltonian accounting for interaction with the substrate. In the next two subsections we derive spectra for the in-plane (\ref{TheoryLATA}) and flexural (\ref{TheoryZA}) phonons within the perturbation theory approach. In Subsection \ref{TheoryMD} we describe the MD simulation, performed to estimate energetic parameters, required by the developed perturbation theory. Section \ref{Results} aggregates results for spectra, phonon lifetimes and heat conductivity. In subsection \ref{ResultsZA} we justify localization of ZA phonons. In subsection \ref{TheoryBTE} we discuss the mechanisms of phonon damping in graphene. The lifetime estimations for in-plane phonons are given in subsection \ref{ResultsLATA} and results for supported graphene heat conductivity are presented and discussed in Subsection \ref{ResultsSummary}.

\section{Theory}
\subsection{Perturbations to conventional graphene Hamiltonian due to the substrate }
\label{Theorygeneral}
We start with considering static properties of graphene on the substrate. Due to the interplay between carbon-carbon and carbon-substrate forces carbon atoms in graphene layer are shifted from their regular positions. These static substrate-induced displacements are described by the vector set $({\bf r}_{l0}, z_{l0})$, where the $l$ index spans all atoms, $\mathbf{r}_{l0}$ and $z_{l0}$ are in-plane and out of plane displacements correspondingly. Thus, we write the following equation for the graphene potential energy
\begin{equation}
  E=\sum_l \left( U_{sub}(\mathbf{r}_{l0},z_{l0}) + \sum_j \frac{K (\delta R_{lj})^2}{2}\right),
\end{equation}
where $U_{sub}$ is the potential energy stemming from the interaction with the substrate, $K$ is first-neighbor interatomic force constant and $\delta R_{lj}$ is change in distance between $l$-th atom and its neighbors. Here and below index $j=1,2,3$ spans the neighbors of atom. After simple calculations we get
\begin{multline}
 E=\sum_l \left( U_{sub}(\mathbf{r}_{l0},z_{l0}) + \sum_j \frac{K}{2} \left[ ({\bf e}_{lj}(\mathbf{r}_{l0}-\mathbf{r}_{j0}))^2 \right. \right. \\ + \left. \left. \frac{((z_{l0}-z_{j0})^2+(r^\tau_{l0}-r^\tau_{j0})^2)^2}{4R^2}\right] \right),
\end{multline}
Here ${\bf e}_{lj}$ is a unit vector along the direction from atom $l$ to its neighbor $j$, $r^\tau$ is a perpendicular to bond in-plane displacement, explicitly $r^\tau_{l0}-r^\tau_{j0}=|(\mathbf{r}_{l0}-\mathbf{r}_{j0})-(\mathbf{e}_{lj}\cdot(\mathbf{r}_{l0}-\mathbf{r}_{j0}))\mathbf{e}_{lj}|$. Positions of carbon atoms on the amorphous substrate satisfy standard equations for classical energy minimum ($\partial E/ \partial \mathbf{R}_l=0$). Introducing small deviations $x_l, y_l, z_l$ from equilibrium positions we get a perturbation of the conventional in-plane phonon Hamiltonian:
\begin{multline}
  \delta H_{in-plane} = \sum_l \left[ \frac{1}{2}\frac{\partial^2 U_{sub}}{\partial x^2}({\bf r}_{l0}, z_{i0}) x^2_l \right. \\ + \frac{1}{2}\frac{\partial^2 U_{sub}}{\partial y^2}({\bf r}_{l0}, z_{i0}) y^2_l \\ \left. + \sum_j \frac{K(r^\tau_{l}-r^\tau_{j})^2}{4R^2}(3(r^\tau_{l0}-r^\tau_{j0})^2+(z_{l0}-z_{j0})^2)\right].
  \label{pert_la}
\end{multline}
Also we get the following perturbation for ZA phonons:
\begin{multline}
  \delta H_{ZA}=\sum_l \left[ \frac{1}{2}\frac{\partial^2 U_{sub}}{\partial z^2}({\bf r}_{l0}, z_{l0}) z^2_l \right. \\ + \left. \sum_j \frac{K (z_l-z_j)^2} {4 R^2} (3(z_{l0}-z_{j0})^2+(r^\tau_{l0}-r^\tau_{j0})^2) \right],
  \label{energyZA}
\end{multline}
The MD simulation described below allows to obtain the equilibrium structure of graphene on the substrate for temperature $T=0$\ K and the corresponding values of atomic displacements ${\bf r}_{l0}$ and $z_{l0}$. The simulation shows that here the terms with $z_{l0}$ are much larger than ones with $r^\tau_{l0}$. An extra difference in prefactor 3 allows to omit the terms with $r^\tau_{l0}$ here, which is reflected in the definition of parameters $\beta_{lj}$ in the beginning of section \ref{TheoryZA}.

\subsection{Theory for LA and TA phonons}
\label{TheoryLATA}
The perturbation Hamiltonian for the in-plane phonons has the following form (cf. Eq. \eqref{pert_la}):
\begin{equation}
  \mathcal{V}=\sum_l \left[\frac{\gamma_l r^2_l}{2}+ \sum_j \frac{\xi_{lj} (r^\tau_{l}-r^\tau_{j})^2}{2} \right].
  \label{pert1}
\end{equation}
Parameters $\gamma_l$ and $\xi_{lj}$ are following:
\begin{eqnarray}
  \gamma_l= \frac{\partial^2 U_{sub}}{\partial x^2}({\bf r}_{l0}, z_{l0}) + \frac{\partial^2 U_{sub}}{\partial y^2}({\bf r}_{l0}, z_{l0}), \\
  \xi_{lj}=\frac{K(3(r^\tau_{l0}-r^\tau_{j0})^2+(z_{l0}-z_{j0})^2) }{4R^2},
\end{eqnarray}
were determined using MD simulations. Let
\begin{eqnarray}
  \label{gammadefinition}
  \gamma=\langle \gamma_l\rangle,  \nonumber \\
  \label{g_def}
  \tilde{\gamma}_l=\gamma_l-\gamma, \\
  \xi=\langle \xi_{lj} \rangle, \nonumber \\
  \tilde{\xi}_{lj}=\xi_{lj} -\xi,
\end{eqnarray}
where angle brackets denote the averaging over disorder configurations. Obviously $\langle \tilde{\gamma}_l \rangle,\langle \tilde{\xi}_{lj} \rangle =0$. So we can divide perturbation \eqref{pert1} in two parts:
\begin{eqnarray}
  \mathcal{V}=\mathcal{V}_1+\mathcal{V}_2, \\
  \mathcal{V}_1=\gamma \sum_l \left[\frac{r^2_д}{2}+\xi  \sum_j \frac{ (r^\tau_{д}-r^\tau_{о})^2}{2}\right], \\
  \label{pert2}
  \mathcal{V}_2=\sum_l \left[\frac{\tilde{\gamma}_l r^2_l}{2}+ \sum_j \frac{\tilde{\xi}_{lj} (r^\tau_{l}-r^\tau_{j})^2}{2} \right].
\end{eqnarray}
It is easy to include $\mathcal{V}_1$ to the exact phonon spectrum, because it only shifts system vibrational eigenvalues ($\omega^2$) by $\gamma$ and slightly renormalizes sound speed $c$. One can see that the term with $\xi$ in $V_1$ contains additional factor $(\mathbf{e}_k-(\mathbf{e}_k\cdot\mathbf{e}_\nu)\mathbf{e}_\nu)^2$, where $\nu=1,2,3$ indicates the bond index and phonon wave wave vector direction $\mathbf{e}_{\mathbf{k}}=\frac{\mathbf{k}}{k}$. At small momenta $k$ its average over momentum angle is almost $1/4$, with negligible trigonal warping. Thus, in Debye model bare phonon spectrum reads:
\begin{equation}
  \omega_k=\sqrt{\frac{\gamma}{m}+\left(1+ \frac{\xi}{2K}\right)c^2 k^2},
  \label{SpectrumLA0}
\end{equation}
where $m$ is carbon atom mass. From this expression one can see that the bare spectrum is gapped. Let $\tilde{c}$ be the renormalized sound speed,
\begin{equation}
  \tilde{c}=c\sqrt{\left(1+ \frac{\xi}{2K}\right)}.
\end{equation}

We use the conventional quantized atom displacement representation:
\begin{equation}
  r_l({\bf R}_l,t)=\frac{1}{\sqrt{2mN}} \sum_k \frac{{\bf p}_k}{\sqrt{\omega_k}} \left( b_k e^{i {\bf k} {\bf R}_l} + b^+_k e^{ - i {\bf k} {\bf R}_l}\right),
  \label{displ_rep}
\end{equation}
where $N$ is the number of unit cells in the system, $b_k$ and $b^+_k$ are bosonic operators, and ${\bf p}_k$ is phonon polarization. For LA phonons ${\bf p}_k ={\bf e}_k$ and for TA phonons ${\bf p}_k$ is perpendicular to ${\bf e}_k$ and lies in graphene plane. Using this equation it is easy to show that
\begin{equation}
  \mathcal{H}_0=\sum_k \omega_k \left( b^+_k b_k + \frac{1}{2}\right),
\end{equation}
as in the case of standard gapless acoustic phonons. Using Eq.\eqref{displ_rep} we have the first part of perturbation \eqref{pert2} in the following form:
\begin{multline}
  \mathcal{V}_2=\frac{1}{4mN} \sum_l \sum_{k_1,k_2} \frac{{\bf p}_{k_1}\cdot{\bf p}_{k_2}\tilde{\gamma_l}}{\sqrt{\omega_{k_1}\omega_{k_2}}} \\ \times \left( b_{k_1}b_{k_2} e^{i ({\bf k}_1 +{\bf k}_2) {\bf R}_l} + b^+_{k_1} b_{k_2} e^{i ({\bf k}_2 -{\bf k}_1) {\bf R}_l} \right) + H.C.,
  \label{pert3}
\end{multline}
where ``H.C.'' is for Hermitian conjugate. Simple estimation shows that main impact to phonon spectrum stems from normal terms in the perturbation and anomalous terms can be omitted. So including the second part of perturbation \eqref{pert2} yields:
\begin{multline}
   \mathcal{V}_2=\frac{1}{2mN} \sum_l \sum_{k_1,k_2} \frac{b^+_{k_1} b_{k_2} e^{i ({\bf k}_2 -{\bf k}_1) {\bf R}_l}}{\sqrt{\omega_{k_1}\omega_{k_2}}} \bigg[ \tilde{\gamma_l}({\bf p}_{k_1}\cdot{\bf p}_{k_2})  \\ +\sum_j \left[ \tilde{\xi}_{lj}(\mathbf{p}_{k_1}-(\mathbf{p}_{k_1}\cdot\mathbf{e}_{lj})\mathbf{e}_{lj})\cdot(\mathbf{p}_{k_2}-(\mathbf{p}_{k_2}\cdot\mathbf{e}_{lj})\mathbf{e}_{lj})  \right.    \\ \left. \times \left( 1- e^{-i {\bf k_1} ({\bf R}_j -{\bf R}_l)}\right)\left( 1- e^{i {\bf k_2} ({\bf R}_j -{\bf R}_l)} \right) \right] \bigg] .\label{pert_la2}
\end{multline}

\begin{figure}
  \noindent
\subfigure{\label{fig1ab} \includegraphics[scale=0.6]{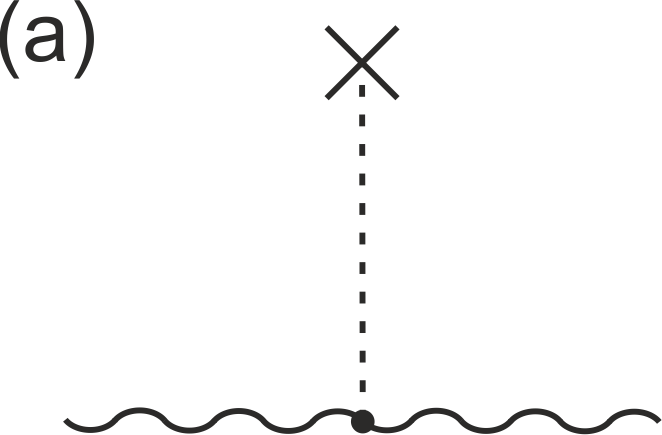}}
\subfigure{\label{fig1ab} \includegraphics[scale=0.6]{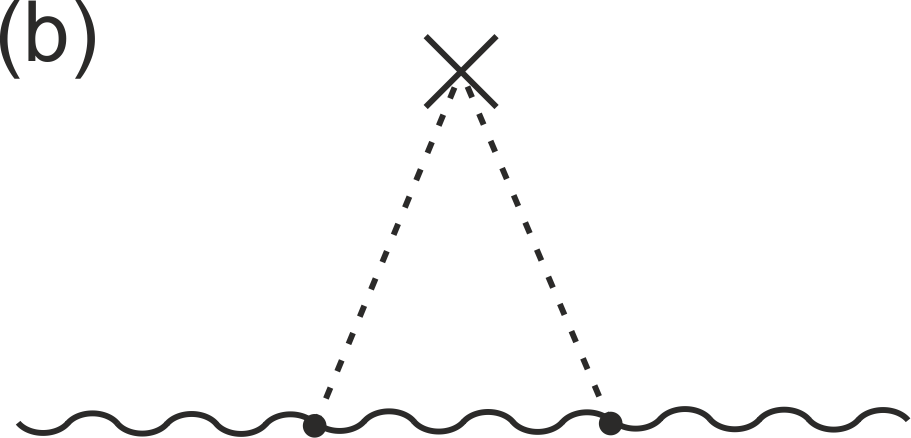}}
\subfigure{\label{fig1c} \includegraphics[scale=0.6]{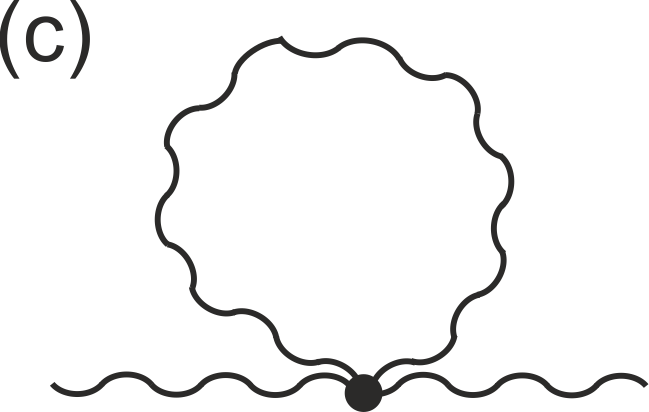}}
  \caption{Diagrams giving corrections to the phonon spectrum. a) - first order in  disorder strength correction, which gives zero  due to disorder parameters definitions (see e.g. Eq.\eqref{g_def}) , b) - second order correction, which is nonzero when the scattering is taking place on the same defect. Waved line is for phonons Green's function and dashed line is for perturbing potential. c) Correction to ZA phonon spectrum due to phonon-phonon interaction.}
\end{figure}

In the following calculations we use this definition for phonon Green's function:
\begin{equation}
  D_0(\omega, {\bf k}) = \frac{2 \omega_k}{\omega^2-\omega^2_k+i0}.
  \label{Green0}
\end{equation}

First order in disorder strength correction to phonon self-energy part is given by the following equation (see. Fig.\ref{fig1ab}):
\begin{multline}
  \Sigma^{(1)}(\omega,{\bf k})=\frac{1}{2m \omega_k} \bigg[ \frac{1}{N}\sum_l \Big( \tilde{\gamma_l} \\ + \sum_j 2\tilde{\xi}_{lj} (1-\cos{\mathbf{k}\cdot\mathbf{e}_{lj}})(\mathbf{p}_{k}-(\mathbf{p}_{k}\cdot\mathbf{e}_{lj})\mathbf{e}_{lj})^2)  \Big) \bigg]=0,
\end{multline}
because expression in the brackets contains $\langle \tilde{\gamma}_l \rangle$ and $\langle \tilde{\xi}_{lj} \rangle$, which are zero. Second order correction is given by the diagram shown in Fig.\ref{fig1ab}. Let us first consider only the term with $\tilde{\gamma}_l$, as the other terms are negligible at small momenta and have no divergences. The corresponding equation reads:
\begin{multline}
  \Sigma^{(2)}(\omega, {\bf k})= \frac{1}{4m^2N^2\omega_k} \\ \times \sum_{l,j} \tilde{\gamma}_l \tilde{\gamma}_j \sum_q \frac{({\bf p}_k \cdot {\bf p}_q)^2 D_0(\omega, {\bf k})}{\omega_q} e^{i ({\bf k} -{\bf q}) ({\bf R}_l - {\bf R}_j)}.
\end{multline}
After averaging over disorder one gets:
\begin{equation}
  \Sigma^{(2)}(\omega, {\bf k})=\frac{\langle \tilde{\gamma}^2_l \rangle v_0}{2m^2 \omega_k} \int \frac{d^2 q}{(2 \pi)^2} \frac{({\bf p}_k \cdot {\bf p}_q)^2 }{\omega^2-\omega^2_q+i0},
\end{equation}
where $v_0$ is the unit cell area. To derive spectrum correction one should put $\omega = \omega_k$ from Eq. \eqref{SpectrumLA0} and integrate
\begin{multline}
  \int \frac{d^2 q}{(2 \pi)^2} \frac{({\bf p }_k \cdot{\bf p}_q)^2}{k^2-q^2+i0} =\frac{1}{4 \pi} \int^{k_D}_{0} \frac{q dq}{k^2-q^2+i0} \\ \approx \frac{1}{8\pi} \left( \ln{\frac{k^2}{k^2_D-k^2}}- i \pi\right),
\end{multline}
where $k_D$ is Debye wave vector. In our dimensionless notations it is equal to $\pi$. Thus the self-energy reads:
\begin{equation}
  \Sigma^{(2)}_\gamma(\omega_k, {\bf k})=\frac{\langle \tilde{\gamma}^2_l \rangle v_0 }{16 \pi m^2 \tilde{c}^2 \omega_k}\left( \ln{\frac{k^2}{k^2_D-k^2}}- i \pi\right).
\end{equation}
Renormalized spectrum can be found from the following equation:
\begin{equation}
  \frac{1}{D(\omega,{\bf k})}=\frac{1}{D_0(\omega,{\bf k})} - \Sigma^{(2)}(\omega_k,{\bf k})=0,
\end{equation}
which has the following solution:
\begin{eqnarray}
  \Omega_k=\omega_k \sqrt{1+\frac{\langle \tilde{\gamma}^2_l \rangle v_0 }{8 \pi m^2 \tilde{c}^2 \omega^2_k}\left( \ln{\frac{k^2}{k^2_D-k^2}}- i \pi\right)}.
\end{eqnarray}
The logarithmic divergencies of this spectrum show that for large enough graphene sheets of the size $L \gg 0.1\,$mm phonons at the bottom and at the top of the band are localized due to the scattering on disorder. But for actual sizes of graphene sheets this is not the case.

Also, there are corrections to phonon spectrum from perturbation \eqref{pert_la2}, which are negligible at small momenta, but can play significant role out of the long wavelenght region. MD analysis shows that the most important correction is the one containing $\tilde{\xi}^2_{lj}$. The corresponding equation reads:
\begin{multline}
\label{sigma2LATAxiterm}
  \Sigma^{(2)}_{\xi}(\omega_k, {\bf k})=\sum_j \frac{32 \langle \tilde{\xi}^2_{lj} \rangle v_0 \sin^2{\frac{k_j}{2}}}{m^2 \tilde{c}^2\omega_k} \\ \times \int \frac{d^2 q}{(2 \pi)^2} \frac{((\mathbf{p}_{k}-(\mathbf{p}_{k}\cdot\mathbf{e}_{j})\mathbf{e}_{j})\cdot(\mathbf{p}_{q}-(\mathbf{p}_{q}\cdot\mathbf{e}_{j})\mathbf{e}_{j}))^2\sin^2{\frac{q_j}{2}}}{k^2-q^2+i0},
\end{multline}
where $k_j$ and $q_j$ are projections of wave vectors $\mathbf{k}$ and $\mathbf{q}$ on the bond direction $\mathbf{e}_j$ with $j$-th atom. So for the in-plane phonons self-energy part of the scattering on disorder has the following form:
\begin{equation}
  \Sigma^{(2)}_{LA}({\bf k})= \Sigma^{(2)}_{\gamma}(\omega_k, {\bf k})+ \Sigma^{(2)}_{\xi}(\omega_k, {\bf k}),
\end{equation}
and the equation for phonon spectrum valid in the whole Brillouin zone reads:
\begin{eqnarray}
  \Omega_k=\omega_k \sqrt{1+2\Sigma^{(2)}_{LA}({\bf k})/\omega_k}.
  \label{LATAspectrumfinal}
\end{eqnarray}

The phonon lifetime due to scattering by the substrate-induced disorder can be extracted from the spectra as its imaginary part $\tau_{\mathrm{substr}}^{-1}(k)=\Im {\Omega_k}$. The only difference in derivations between LA and TA phonons is in polarizations in Eq. \eqref{sigma2LATAxiterm}.

\subsection{Theory for ZA phonons}
\label{TheoryZA}
For ZA phonons one can see from expression \eqref{energyZA} that we have two different perturbation parts. So let us introduce two sets of parameters, which can be calculated from MD simulation:
\begin{eqnarray}
  \alpha=\left\langle \frac{\partial^2 U_{sub}}{\partial z^2}({\bf r}_{l0}, z_{l0}) \right\rangle, \\
  \alpha_l=\frac{\partial^2 U_{sub}}{\partial z^2}({\bf r}_{l0}, z_{l0}) - \alpha, \\
  \label{betadefinition}
  \beta=\left\langle \frac{3 K} {4 R^2} (z_{l0}-z_{j0})^2 \right\rangle, \\
  \beta_{lj}=\frac{3 K} {4 R^2} (z_{l0}-z_{j0})^2-\beta.
\end{eqnarray}
Obviously $\langle \alpha_l\rangle=\langle \beta_{lj}\rangle=0$. Regular translationally invariant part of the Hamiltonian has the following form:
\begin{equation}
\label{H0ZA}
  \mathcal{H}^{(0)}_{ZA}=\sum_i \left[ \frac{p^2_{zi}}{2m} + \frac{\alpha z^2_i}{2} + \sum_j \frac{\beta (z_i-z_j)^2}{2}\right].
\end{equation}
Comparing this Hamiltonian with the conventional one for suspended graphene one can see that spectrum is the following:
\begin{equation}
\label{spectrum35}
  \omega_k=\sqrt{\frac{\alpha}{m}+c^2_{ZA}k^2},
\end{equation}
where
\begin{equation}
  c_{ZA}=\left(\frac{2\beta}{K}\right)^{1/2} c.
  \label{cZA}
\end{equation}

It is instructive to compare the obtained dispersion of flexural phonons with one from paper by Amorim and Guinea\cite{PhysRevB.88.115418}. First, both dispersions are gapped with gap width controlled by strength of graphene-substrate interaction $\alpha$, which defines the minimal vibration energy in the field of the substrate (cf. with $g$ parameter from Ref. \cite{PhysRevB.88.115418}). The difference is in the power of the phonon wave vector $q$. Amorim and Guinea consider the quadratic in displacement intrinsic bending rigidity of graphene for restoring force and as a result the dispersion is quadratic for large $q$. On the contrary, as one sees from Eq. \eqref{H0ZA}, here the quadratic in displacement force associated with $\beta$ terms plays a role of restoring force. It results in linear bare specrtrum of flexural phonons for graphene on substrate for large $q$, which significantly differs from the spectrum of freestanding graphene and graphene on crystalline substrate. In fact, the dispersion of ZA phonons of graphene on amorphous substrate has rather extrinsic than intrinsic origin.

As the phonon-phonon interaction is significant in the case of ZA phonons, we consider the anharmonic term in the Hamiltonian,
\begin{equation}
  V=U\sum_{l,j} (z_l-z_j)^4,
\end{equation}
where summation is conducted over all nearest neighbors, and $U=K/8R^2$. In quantized form:
\begin{multline}
  V=\frac{3U}{2m^2N}\sum_{k_1+k_2=k_3+k_4}\frac{b^+_{k_1}b^+_{k_2}b_{k_3}b_{k_4}}{\sqrt{\omega_{k_1}\omega_{k_2}\omega_{k_3}\omega_{k_4}}} \\ \times \sum_j \left(1-e^{-i k_{1j}} \right)  \left(1-e^{-i k_{2j}} \right) \left(1-e^{i k_{3j}} \right)  \left(1-e^{i k_{4j}} \right).
\end{multline}
$k_{1j},k_{2j},k_{3j},k_{4j}$ are the projections of wave vectors $\mathbf{k}_1,\mathbf{k}_2,\mathbf{k}_3,\mathbf{k}_4$ on the direction from arbitrary atom to its $j$-th neighbor in the real space. Linear in $U$ correction is given by the diagram in Fig. \ref{fig1c} with corresponding equation in Matsubara technique writing:
\begin{equation}
  \Sigma_T(\mathbf{k})=\frac{12K v_0}{2mR^2\omega_k} \sum_j \sin^2{\frac{k_j}{2}} \int \frac{d^2 q}{(2 \pi)^2} \frac{\sin^2{\frac{q_j}{2}}}{\omega_q} \coth{\frac{\omega_q}{2T}}.
\end{equation}

Thus, the spectrum used henceforth is
\begin{equation}
  \label{specZA}
  \omega^T_k=\omega_k \sqrt{1+\frac{2\Sigma_T(\mathbf{k})}{\omega_k}}
\end{equation}

Perturbation Hamiltonian has the following form:
\begin{equation}
  \label{pertZA}
  \mathcal{V}=\sum_l \left[ \frac{\alpha_l z^2_l}{2} +\sum_j \frac{\beta_{lj} (z_l-z_j)^2}{2} \right],
\end{equation}
once again we omit anomalous terms and rewrite the perturbation:
\begin{multline}
  \mathcal{V}=\frac{1}{2mN}  \sum_l \sum_{k_1,k_2} \frac{ b^+_{k_1} b_{k_2}}{\sqrt{\omega_{k_1}\omega_{k_2}}} e^{i ({\bf k}_2 -{\bf k}_1){\bf R}_l} \Bigl[ \alpha_l \\  + \sum_j \beta_{lj} \left( 1- e^{-i {\bf k_1} ({\bf R}_j -{\bf R}_l)}\right)\left( 1- e^{i {\bf k_2} ({\bf R}_j -{\bf R}_l)} \right) \Bigr]
\end{multline}
At the small momenta $k \ll 1$ the second order spectrum correction is given only by the term with $\alpha_l$. As in the previous subsection this correction is logarithmically divergent for long-wavelength phonons. This indicates their localization at $k \ll a_0/L$, where $a_0$ is the lattice parameter and $L$ is graphene flake size. Expressions for all corrections of the second order in disorder strength are presented in Appendix \ref{append}. Based on results of MD simulations presented in Subsec. \ref{TheoryMD}, we only keep the two main terms producing the highest contributions, which give the following self-energy:
\begin{equation}
  \Sigma^{(2)}_{ZA}({\bf k})=\frac{v_0}{m^2 \omega^T_k} \left(  \langle \alpha^2_l \rangle I_1(k) + 32\langle \beta^2_{lj}\rangle \sum_j  \sin^2{\frac{k_j}{2}} I_{2j}({\bf k}) \right),
\end{equation}
where
\begin{eqnarray}
  I_1({\bf k})=\frac{1}{2}\int \frac{d^2 q}{(2 \pi)^2}\frac{1}{(\omega^T_k)^2-(\omega^T_q)^2+i0}, \\
  I_{2j}({\bf k})=\int \frac{d^2 q}{(2 \pi)^2} \frac{\sin^2{\frac{q_j}{2}}}{(\omega^T_k)^2-(\omega^T_q)^2+i0}.
\end{eqnarray}
The corresponding expression for ZA phonon spectrum reads:
\begin{eqnarray}
\label{ZAspectrumfinal}
  \Omega_k=\omega^T_k \sqrt{1+2\Sigma^{(2)}_{ZA}({\bf k})/\omega^T_k}.
\end{eqnarray}

\subsection{Molecular dynamics simulation}
\label{TheoryMD}
The GROMACS\cite{gromacs} package was used to perform all MD simulations. The 72\AA\,x 72\AA\,x 36\AA\,amorphous SiO$_2$ substrate consisting of 12000 atoms was rigid. The round graphene sheet of diameter $D \approx 40$\,\AA\ consisted of 481 carbon atoms. The C-C interactions within the graphene sheet were modeled with harmonic  potential $K = 27$\,eV/\AA$^2$, which corresponds to the first neighbor interatomic force constant from \cite{dubayPhysRevB.67.035401}. This model is consistent with employed analytical description of graphene vibrational properties. Interactions between C atoms of graphene and Si and O atoms of the substrate were modeled with Lennard-Jones potential with parameters taken from \cite{MDgraphene}.

Graphene sheet initial position was 1\,\AA\ above the substrate. Simulation was performed with 0.02\,ps timestep, the temperature was controlled with weak coupling algorithm\cite{bussi2007canonical}. During the first 500\,ps graphene sheet was attracted by the substrate and started planar diffusion on its surface. Then from initial value of 300\,K temperature was lowered to zero for 500\,ps using velocity rescale temperature coupling \cite{bussi2007canonical}.

\begin{figure}
\centering
\includegraphics[width=0.5\textwidth]{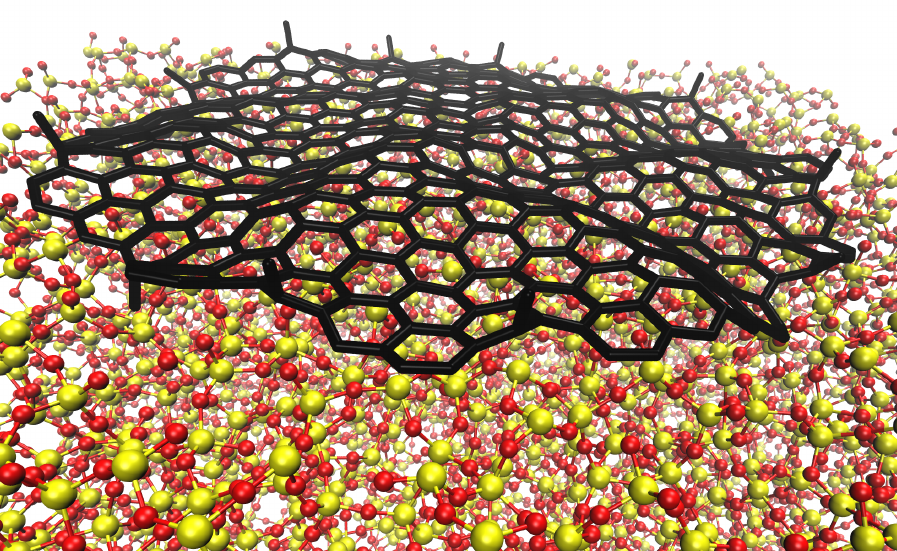}
\caption{Color online. Final structure of graphene on amorphous SiO$_2$ obtanied with MD simulation.}
\label{figMD}
\end{figure}

The obtained freezed equilibrium configuration of graphene on SiO$_2$ substrate was treated in Mathematica \cite{ram2010} package to derive the parameters required by the theory. Only carbon atoms with three neighbors were considered. The results exhibit no significant dependence on initial conditions of MD simulations. The required by perturbation theory graphene-substrate interaction energetic parameters are listed in the Table I. Fig. \ref{figMD} shows the obtained geometry of graphene on the amorphous SiO$_2$ substrate. The obtained value of $\beta=0.09$\ eV/\AA$^2$ corresponds to the graphene-substrate coupling parameter $g=5 \times 10^{19}$\ J/m$^4$, which is 4 times smaller than estimation from Ref. \cite{0295-5075-91-5-56001}. This is due to not complete slippage between graphene sheet and rough amorphous SiO$_2$ substrate.

The atomic Z coordinate root mean square displacement for graphene on the substrate with respect to initial unperturbed graphene is 0.04\,nm. For the X and Y coordinates we obtain 0.007\,nm. With high accuracy $\beta=3\xi$ and $\sqrt{<\beta_{ij}^2>} = 3\sqrt{<\xi_{jl}^2>}$.

\begin{table*}[!ht]
\centering
\caption{The obtained with MD parameters required by perturbation theory. All values are given in eV/\AA$^2$}
\begin{tabular}{ l c c c c c c r }
\hline
\hline
\multicolumn{4}{c}{In-plane phonons} & \multicolumn{4}{c}{ZA phonons}\\
$\gamma$& $\sqrt{<\tilde{\gamma}_l^2>}$ & $\xi$ & $\sqrt{<\xi_{lj}^2>}$ & $\alpha$ & $\sqrt{<\alpha_l^2>}$ & $\beta$ & $\sqrt{<\beta_{lj}^2>}$ \\
\hline
0.006 & 0.015 & 0.2 & 0.28 & 0.09 & 0.11 & 0.59 & 0.82  \\
\hline
\hline
\end{tabular}
\end{table*}

\section{Results and discussion}
\label{Results}

\subsection{Spectra and relaxation times of ZA phonons}
\label{ResultsZA}
As it was discussed above, the bare specrtum of ZA phonons for graphene on disordered substrate given by Eq. \eqref{spectrum35} differs from the one for freestanding graphene and graphene on crystalline substrate given by $\omega_k=A k^2$, where $A \approx 3.1\cdot10^{-3}$cm$^2$s$^{-1}$ \cite{PhysRevB.68.134305}. The phonon-phonon interaction described by \eqref{specZA} leads to the effective increasing of $c_{ZA}$. For $T=300$\ K the value of $c_{ZA}$ is 6\,km/s or approximately 30\% of the in-plane phonon velocity.

\begin{figure}
\centering
\subfigure{\label{figZAspectrum} \includegraphics[width=0.45\textwidth]{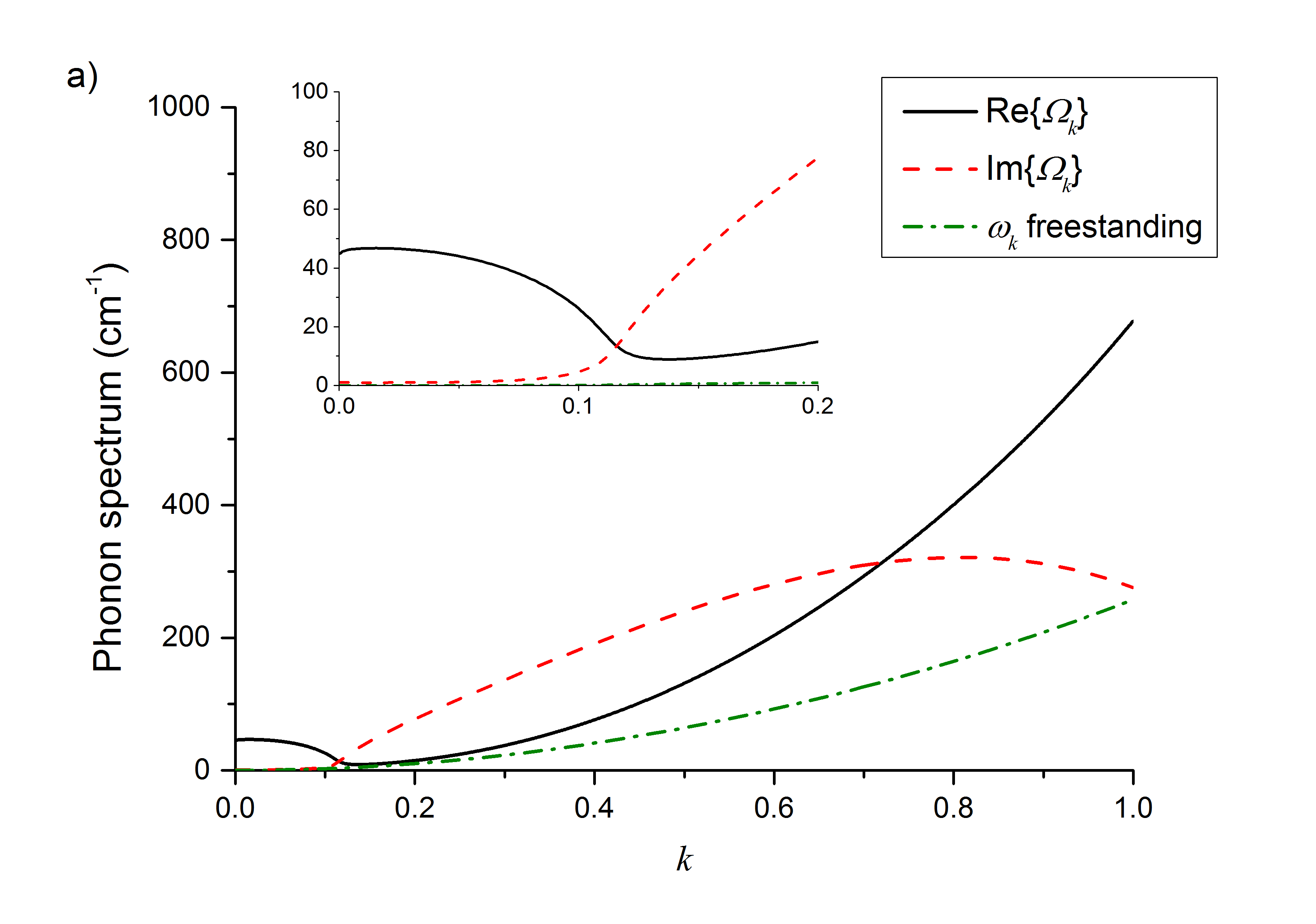}}
\subfigure{\label{figZAImdivRe}\includegraphics[width=0.45\textwidth]{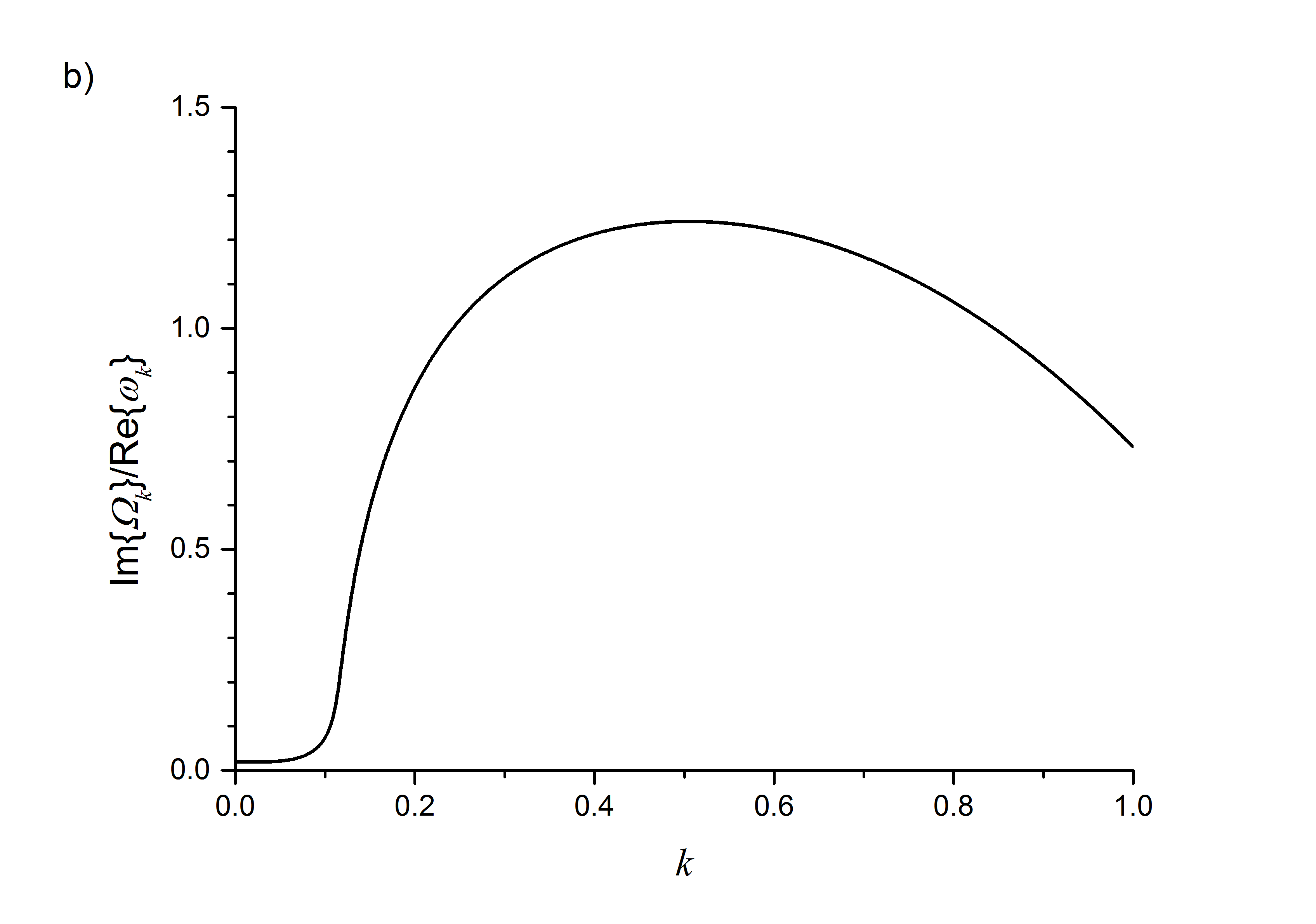}}
\caption{Color online. a) Calculated spectrum of ZA phonons in graphene on amorphous substrate. Solid black curve and red dashed curve are for real and imaginary parts of ZA phonon spectrum at $T=300$\,K for graphene on amorphous substrate, respectively. Green dash dot curve is for ZA phonon spectrum of freestanding graphene. b) Ratio of imaginary part of ZA phonon spectra to real part of bare spectra for graphene on disordered substrate.}
\end{figure}

Fig. \ref{figZAspectrum} shows the real and the imaginary parts of ZA phonon spectrum for $T=300$\ K calculated with Eq. \eqref{ZAspectrumfinal} using parameters from Table I and Fig. \ref{figZAImdivRe} shows the ZA phonon spectrum imaginary to real part ratio.

In the considered system we have strong disorder for ZA phonons with spectrum $\omega^T_k$ calculated via Eq. \eqref{specZA}. It is well known that even disorder with small concentration leads to localization of long-wavelength excitations (for bosonic systems see E.G. Refs. \cite{PhysRevB.81.064301,luckyanova2016phonon,magnons}). But here "impurities concentration" is equal to $1$, providing spectrum corrections of the order of the pure spectrum \eqref{specZA} in the whole Brillouin zone (see Fig. \ref{figZAspectrum}), which makes ZA phonons overdamped (see Fig. \ref{figZAImdivRe}) and their nature is a  question of further considerations, the possible scenario is localization of all ZA phonon modes. Anyway, from our theory it is quite natural to make a conclusion that ZA phonons do not contribute to graphene on amorphous substrate heat conductivity.

\subsection{Phonon scattering mechanisms in graphene}
\label{TheoryBTE}

To understand the impact of substrate on graphene heat conductivity it is necessary to compare the phonon scattering on the substrate-induced disorder with other mechanisms of phonon relaxation in graphene.

The conventional mechanisms of phonon lifetime reduction in graphene are scattering on the graphene flake boundaries and anharmonic three-phonon scattering processes, while point defect scattering is weak compared to other mechanisms. The single mode relaxation time approximation (SMRTA) is always used to describe the boundary scattering and corresponding phonon lifetime can be written as $\tau_L^{-1}=v_k/L$, where $v_k = \frac{\partial \omega_k}{\partial k}$ is the phonon group velocity, $L$ estimates the graphene sample size. For the in-plane phonons the averaged sound velocity can be taken without loss of accuracy.

Currently there is no established theory on the suspended graphene heat conductivity due to complexity of the anharmonic processes. Various approaches including usage of BTE within SMRTA \cite{6235226}, exact iterative solution of BTE accounting for three-phonon scattering \cite{PhysRevB.82.115427,spectralHeatCond} and MD simulations \cite{copper} yield different values of graphene heat conductivity $\kappa$ and distribution of heat flux between LA,TA and ZA modes. The relevant data is accumulated in Table 3 from the review \cite{0953-8984-28-48-483001}.

The contribution of anharmonic processes to phonon damping can be described within the SMRTA with the power law \cite{srivastava1990physics,callaway1959model,6235226}
\begin{equation}
  \tau_{\mathrm{anh}}^{-1}=(B_N+B_U\exp(-\Theta/B T))\omega^2 T^3,
  \label{tauanh}
\end{equation}
where $T$ is temperature, $\omega$ is phonon frequency, $\Theta \approx 1000$\ K is the Debye temperature, $B = 3$, $B_N = B_U/2$ and $B_U = 7.7\cdot10^{-25}$ sK$^{-3}$, see Ref. \cite{6235226}.

Recently it was shown that the approach for considering anharmionic processes in graphene, which reflects all peculiarities of three-phonon scattering should look beyond SMRTA. This problem requires exact solution of BTE for the three-phonon scattering. In this model, the strong mixing of the in-plane and the flexural phonon modes was justified by Linsay et al. \cite{PhysRevB.82.115427}. Authors show that selection rules for the phonon decay include obligate involving of even number of flexural phonons. The direct $LA,TA \rightarrow ZA+ZA$ and inverse processes $ZA+ZA \rightarrow LA,TA$ provide a balance between in-plane and flexural phonon distribution functions. To calculate the value of heat conductivity authors relate an effective relaxation time for each phonon mode from obtained distribution function correction. The contribution of ZA phonons to free standing graphene heat conductivity of about 80\% in wide temperature range was reported. Another result of this study is that SMRTA and exact BTE solution give 2-3 times discrepancy in $\kappa$ magnitude for LA and TA modes (up to 8 times for ZA mode \cite{PhysRevB.82.115427}). Singh et al. in Ref. \cite{spectralHeatCond} use a similar approach and also conclude that ZA phonons give significant contribution to $\kappa$.
The authors give an estimate of the total phonon lifetime, which differs from the result of the SMRTA approach up to 3 orders of magnitude. They also show that the lifetime of the in-plane phonons with respect to conversion to the ZA mode is twice longer than total lifetime.

The following two approaches were used to estimate the value of $\tau_{\mathrm{anh}}$ for deriving graphene heat conductivity $\kappa$ estimation. First, $\tau_{\text{anh}}$ was calculated by formula \eqref{tauanh}, see Ref.\cite{6235226}. Second, it was adopted from Ref. \cite{spectralHeatCond}, see Fig. 6 there.

The interaction with the substrate will affect the selection rules for phonon scattering. The localization of flexural phonons (see section \ref{ResultsZA}) can change the distribution of the heat flux between LA, TA and ZA modes. Without understanding the nature of localized ZA phonons in graphene on amorphous substrate one can not judge about mixing of in-plane and ZA phonons, which opens a field for further investigations.

The total phonon relaxation time was estimated via Matthiessen's rule $\tau_{\mathrm{total}}^{-1}=\tau_{L}^{-1}+\tau_{\mathrm{anh}}^{-1}+\tau_{\mathrm{substr}}^{-1}$. The estimation of contribution to graphene heat conductivity from a given phonon mode can be written as
\begin{equation}
\label{BTE}
  \kappa = \frac{\hbar}{h} \int kdk \tau_{\mathrm{total}}(k) \omega_k \frac{\partial N^0(\omega_k)}{\partial T} v_k^2,
\end{equation}
where $N^{(0)}$ is equilibrium Bose distribution function and $h=0.335$\ nm is the graphene layer thickness.

Athough ZA phonons are localized, calculating ZA phonons contribution to thermal conductivity using Eq. \eqref{BTE}, where one assumes $v_k = \frac{\partial}{\partial k} \Re \{\Omega_k\}$ and $\tau_{\mathrm{substr}}^{-1} = \Im \{\Omega_k\}$, yeilds less than 20\,Wm$^{-1}$K$^{-1}$. The conribution of ZA phonons can de thus anyway neglected due to significant reduction of lifetime. Fig. \ref{figZAspectrum} with the characteristic magnitude of ZA phonon spectrum imaginary part shows that $\tau_{\mathrm{substr}}$ is lower than 0.1\,ps in most volume of Brillouin zone.

\subsection{Spectra and relaxation times of in-plane phonons}
\label{ResultsLATA}
\begin{figure}
\centering
\includegraphics[width=0.5\textwidth]{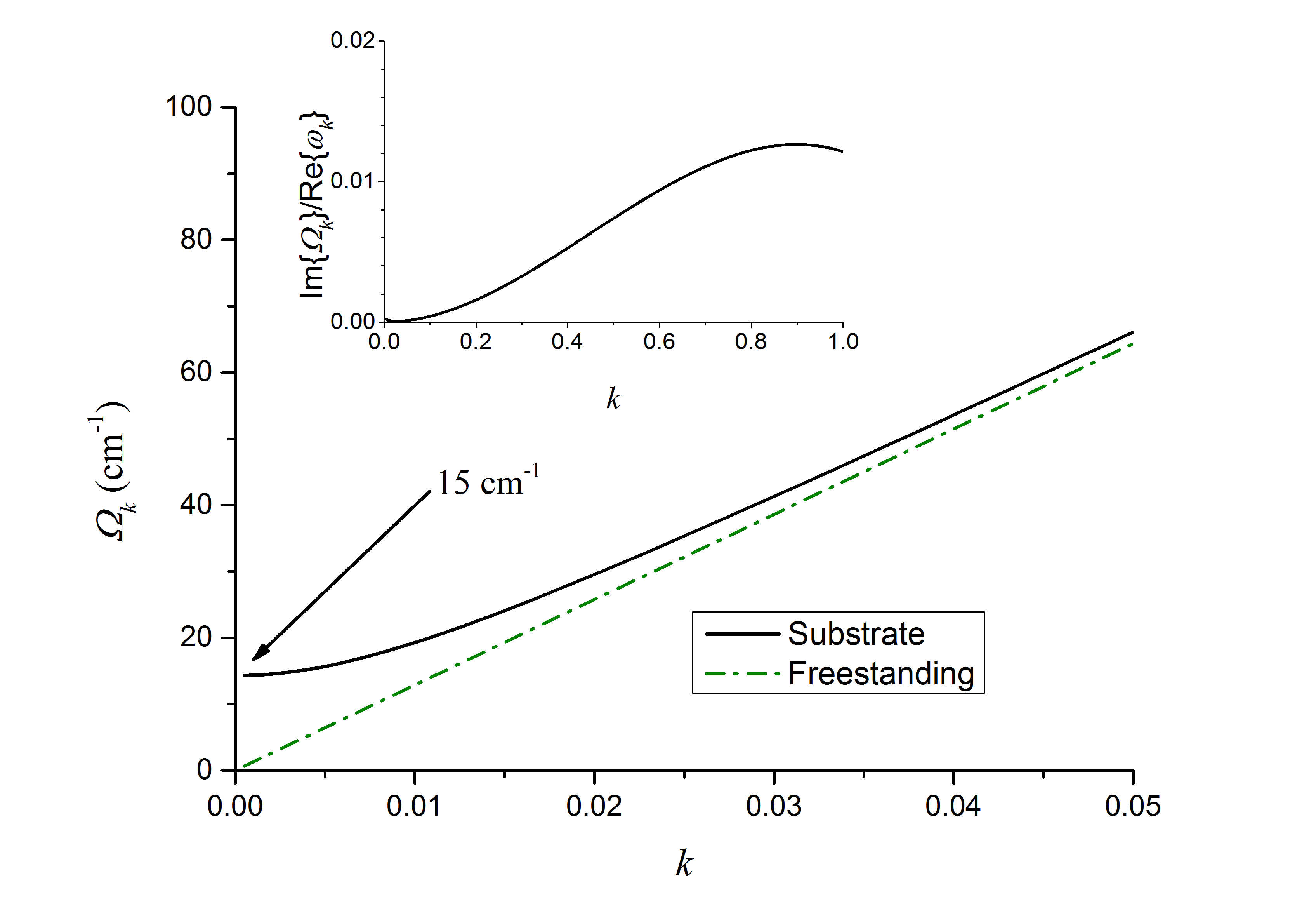}
\caption{Color online. Dispersion relations of in-plane acoustic phonons in freestanding graphene and in graphene on substrate. The inset shows the LA phonon spectrum imaginary to real part ratio for graphene on disordered substrate. Phonon wave vector $k$ is given in units of $\pi/a_0$ and $k = 1$ corresponds to the boundary of the Brillouin zone.}
\label{figLATAspectra}
\end{figure}

Substrate influence on the in-plane phonons in graphene constitutes in the following effects. First, according to Eq. \eqref{SpectrumLA0} interaction with substrate leads to the opening of band gap of approx. 22 K, see Fig. \ref{figLATAspectra}. It results the phonon occupation number suppression for low temperatures and corresponding reduction of graphene thermal conductivity for temperatures lower than 20K.

Secondly, the scattering by the substrate-induced disorder leads to additional reduction of phonon lifetime. The predictions on the strength of this effect strongly depends on the model for $\tau_{\mathrm{anh}}$ employed. So for $\tau_{\mathrm{anh}}$ calculated by Eq. \eqref{tauanh} both in suspended and supported graphene the boundary scattering is the dominant mechanism of long wavelength phonon damping and defines the heat conductivity at low temperatures. At low temperatures only long wavelength phonons are excited and give contribution to heat conductivity. On the contrary for $\tau_{\mathrm{anh}}$ adopted from Ref. \cite{spectralHeatCond} such phonons are damped due to anharmonic processess. For shorter wavelength phonons, which are excited at temperatures above 100\ K, the dominant mechanism corresponds to the anharmonic processess and to substrate-induced disorder scattering.

\begin{figure}
\centering
\includegraphics[width=0.5\textwidth]{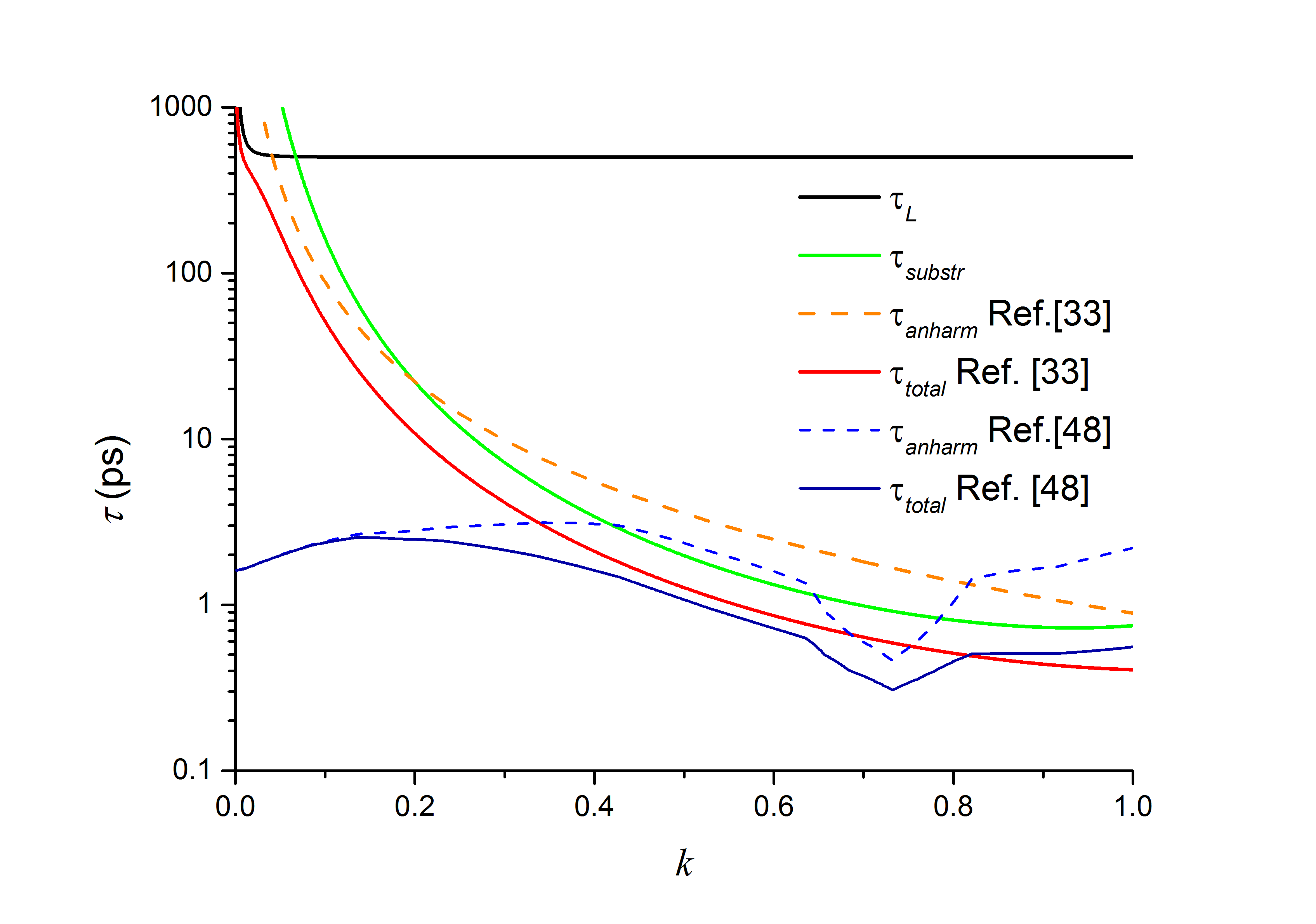}
\caption{Color online. LA phonons relaxation times corresponding to various mechanisms of scattering. Black solid curve is for relaxation time related to boundary scattering $\tau_L$ with $L=10\,\mu$m. Green solid curve is for substrate-induced phonon damping. Orange dashed curve is for anharmonic processes for temperature $T$=300\ K obtained by Eq. \eqref{tauanh} and solid red curve is for total phonon lifetime in this model, derived with Matthiessen's rule. The light blue dashed curve is for $\tau_{\mathrm{anh}}$ at $T$=300\ K adopted from Ref. \cite{spectralHeatCond} and blue solid curve is for corresponding total lifetime.}
\label{figLATAlifetimes}
\end{figure}

Fig. \ref{figLATAlifetimes} shows the dependence of the relaxation time of LA phonons on the phonon wave vector magnitude for various mechanisms. Phonon lifetime for anharmonic processes is given for 300 K.

For TA phonons the effect of substrate is weaker due to polarization effects stemming from integral \eqref{sigma2LATAxiterm} and anharmonic processes are more important than the substrate effect.

The obtained values of the in-plane phonon lifetime yield the phonon drag thermopower at the level of several $\mu$VK$^{-1}$, which is much smaller than diffusion contribution to thermopower \cite{koniakhin2013phonon}.

\subsection{Graphene heat conductivity}
\label{ResultsSummary}

As it was shown above, the graphene-substrate interaction suppresses the contribution of ZA phonons to heat conductivity in the whole range of temperatures for graphene bonded with amorphous SiO$_2$ substrate. This contribution is negligible due to ZA phonon localization and extremely short lifetime. Therefore, the total heat conductivity is governed by in-plane phonons.

The behavior of heat conductivity depends on the applied model for $\tau_{\mathrm{anh}}$ estimation. For the $\tau_{\mathrm{anh}}$ given by Eq. \eqref{tauanh}, at temperatures below 100\,K the in-plane phonon heat conductivity is governed by boundary scattering for both supported and suspended graphene. For $\tau_{\mathrm{anh}}$ adopted from Ref.\cite{spectralHeatCond} the anharmonic processes play main role at temperatures below 100 K. At temperatures above 100\,K the dominant mechanism corresponds to the anharmonic processes and to substrate-induced disorder scattering in both models for supported graphene. As a result the supported graphene heat conductivity is several times smaller than for freestanding graphene, which is in a qualitative agreement with available experimental data.

For $\tau_{\mathrm{anh}}$ adopted from \cite{spectralHeatCond} in the whole temperature range the anharmonic processes give main contribution to LA and TA phonon damping (see Fig. \ref{figLATAlifetimes}), at the same time the flexural phonons contribution in this model reaches 90\%. Thus due to localization of flexural phonons the reduction of supported graphene heat conductivity is drastical.

\begin{figure}
\centering
\includegraphics[width=0.49\textwidth]{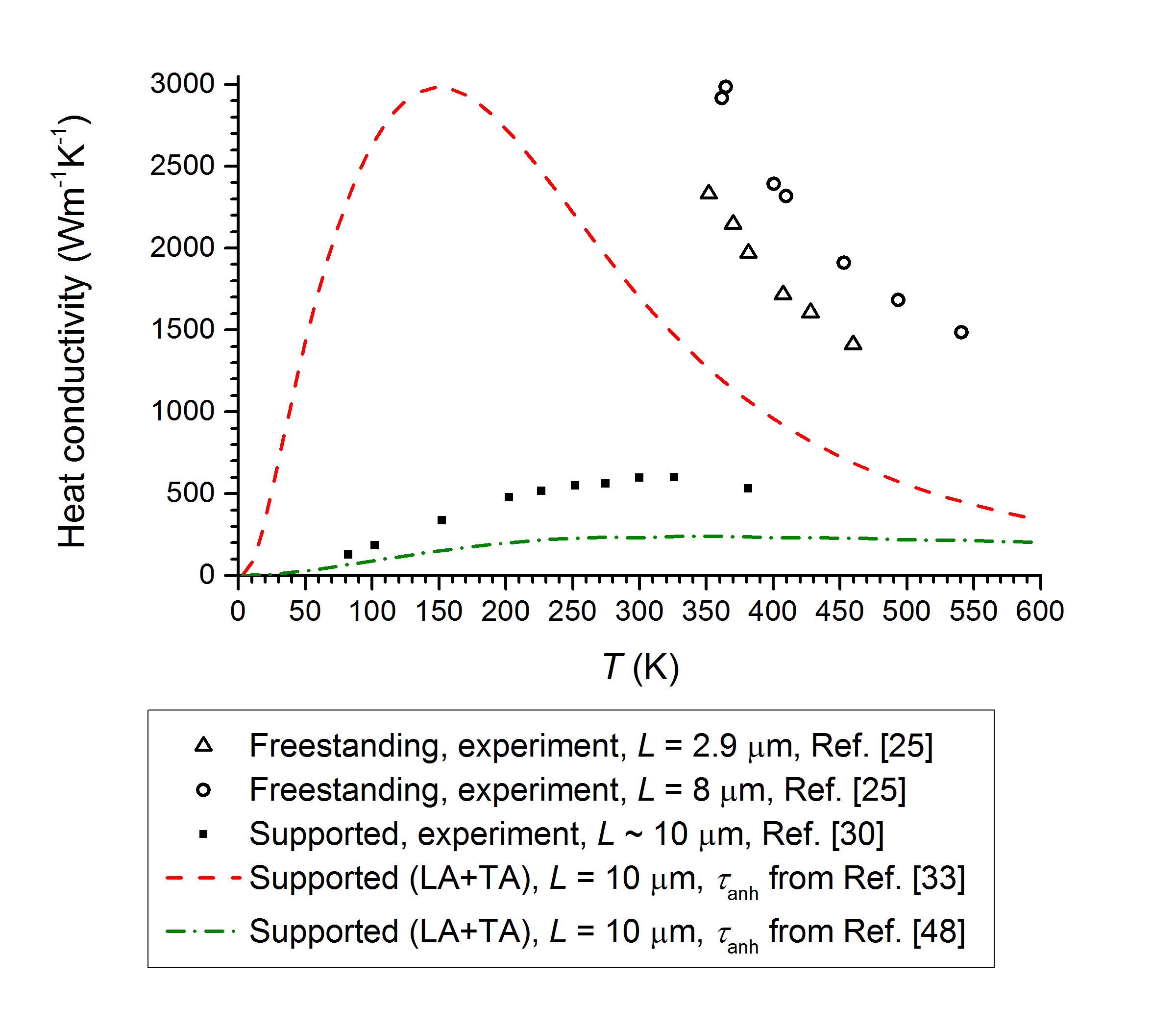}
\caption{Color online. Experimental data on heat conductivity of freestanding graphene for $L=2.9\ \mu$m (open triangles), $L=8\ \mu$m (open circles)\cite{chen2010raman} and heat conductivity of graphene on amorphous substrate (black solid squares)\cite{seol2010two}. Red dashed curve denotes heat conductivity of graphene on amorphous substrate calculated with BTE as a sum of LA and TA phonons contributions with $\tau_{\mathrm{anh}}$ given by Eq. \eqref{tauanh} (see Ref. \cite{6235226}). Green dash-dot curve is for the same but with $\tau_{\mathrm{anh}}$ adopted from Ref. \cite{spectralHeatCond}.}
\label{figLATAheatcond}
\end{figure}

Fig. \ref{figLATAheatcond} shows the experimentally measured heat conductivity of freestanding and supported graphene and corresponding theoretical predictions based on the developed perturbation theory and BTE.

Experimental data from Ref. \cite{seol2010two} indicate significant suppression of the supported graphene heat conductivity below 300\,K and shifting the maximum from approx. 200\,K for freestanding graphene to 300\,K.

\begin{figure}
\centering
\subfigure{\label{fig7a} \includegraphics[width=0.5\textwidth]{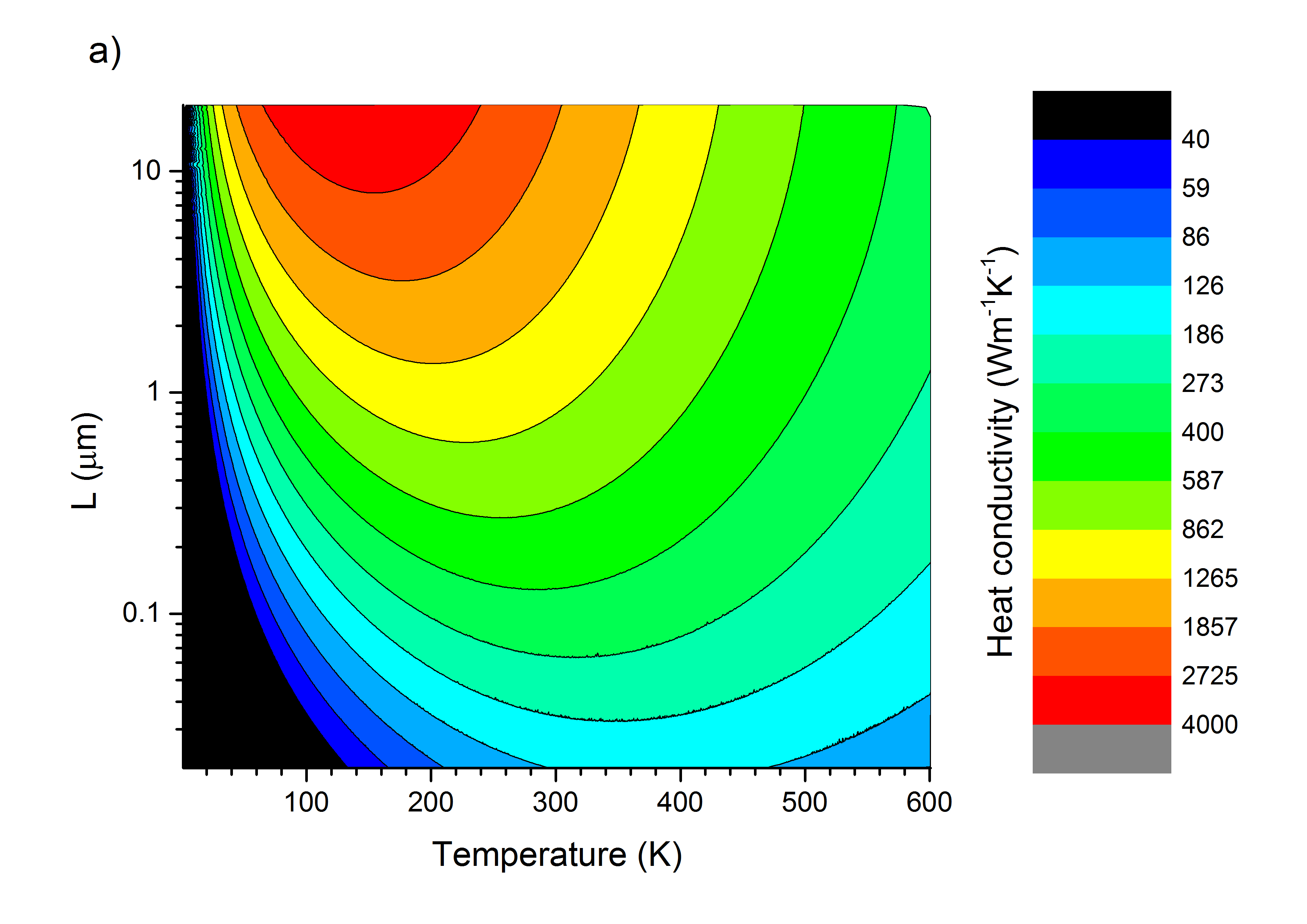}}
\subfigure{\label{fig7b} \includegraphics[width=0.5\textwidth]{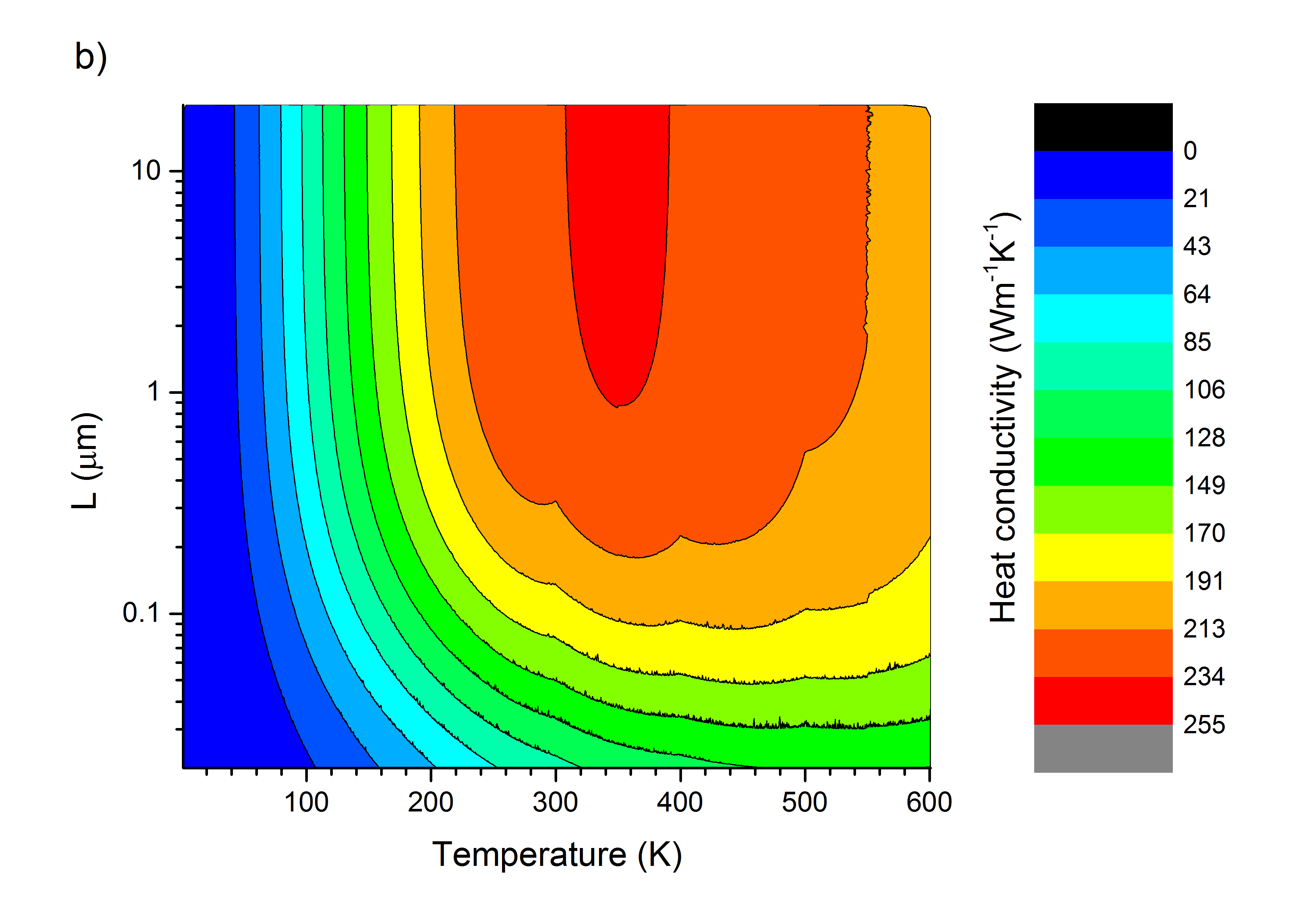}}
\caption{Color online. The obtained within the developed theory dependence of supported graphene heat conductivity on temperature $T$ and size $L$. For panel a) the $\tau_{\mathrm{anh}}$ was estimated with Eq. \eqref{tauanh} (see Ref. \cite{6235226}) and for panel b) $\tau_{\mathrm{anh}}$ was adopted from Ref. \cite{spectralHeatCond}.}
\label{figTandL}
\end{figure}

Using $\tau_{\mathrm{anh}}$ from Eq. \eqref{tauanh} leads to the serious overestimation of heat conductivity, especially at low temperatures. As it can be seen from Fig. \ref{fig7a} to bring theory in agreement with the experiment,  the assumption $L=0.2$ $\mu$m is sufficient, whereas the experiment yields graphene sheet size of several microns.

On the contrary taking $\tau_{\mathrm{anh}}$ adopted from Ref. \cite{spectralHeatCond} leads to twofold underestimation of supported graphene heat conductivity. In this model heat conductivity does not significantly depend on graphene flake size $L$ for $L>0.5$\ $\mu$m, see Fig. \ref{fig7b}.

Alternatively, to explain the observed experimental behavior of heat conductivity, the authors of Ref. \cite{seol2010two} develop a theory, which is based on the Fermi golden rule and apply Klemens formalism\cite{klemens1955scattering} to the case of a large area spot contact between graphene and substrate. In their model, the suppression of long wavelength phonon lifetime stems from an arguable assumption of a constant phonon scattering matrix element. The conventional form of matrix element given in \cite{klemens1955scattering} is quadratic in the phonon frequency, which yields suppression of high frequency phonons instead, see Eq. (32) in Ref. \cite{klemens1994thermal}.

It is also necessary to compare the obtained data on the phonon lifetimes and graphene thermal conductivity with results of the MD simulations. Fig. 2 from Ref. \cite{MDgraphene} shows that the lifetime of both in-plane and ZA phonons in suspended graphene lies in the range from 10 to 25\,ps. For supported graphene, the characteristic lifetimes of all phonon modes are from several picoseconds to 10\,ps. These results are in contradiction with the relatively weak effect of the substrate on the in-plane phonons and dramatic suppression of the ZA phonon lifetime predicted here.

Other studies lie in agreement with our predictions. MD simulations performed in Ref. \cite{ong2011effect} yield $90\%$ reduction of ZA phonon contribution to the thermal conductivity and spectrum deformation for actual strength of the graphene-substrate interaction, which coincides with the results of the present study. The MD study in Ref.\cite{wei2014mode} indicates several times reduction of in-plane phonon lifetime and several orders reduction of ZA phonon lifetime in supported graphene, which qualitatively agrees with present study. Finally, Ref. \cite{zhu2016phonons} predicts localization of excitations in amorphous graphene and corresponding two-fold decrease of heat conductivity.

The predictions on graphene heat conductivity are robust with respect to variation of the energetic parameters from the Table I. Although we account only pair harmonic potential for C-C bonds in MD simulations, we argue that using more sophisticated model (E.G. Tersoff \cite{tersoff1988empirical} and optimized Tersoff \cite{lindsay2010optimized} potentials accounting for three-atomic torsional rigidity and potentials accounting for next neighbors interactions \cite{dubayPhysRevB.67.035401}) would slightly renormalize the parameters and the obtained qualitative picture remains intact.

The employed model lies in agreement with the fact that graphene sheets conform to the underlying silicon oxide substrate, reported in \cite{ishigami2007atomic,geringer2009intrinsic}. However, this model does not account for long-range ($20$\,nm) height correlations of the substrate and graphene. Also these studies report that graphene sheet conforms the substrate corrugations but with smaller amplitude, which is an evidence of graphene being partially suspended. The elucidation of the geometry of graphene interacting via van der Waals force with amorphous substrates is to be clarified in further experimental studies and MD simulations. The developed model is rather related to the atomically smooth substrates. The considered effect of amorphous substrate on graphene intrinsic heat conductivity is important when creating graphene-based electronic devices with the heat-sink functions placed on graphene pathways and should be taken into account when managing circuit thermal parameters.

\begin{acknowledgments}

S.K. acknowledges support by the Russian Science Foundation (Project No. 16-19-00075 “Thermoelectric generator with record parameters based on carbon nanostructures: development of scientific bases"). O.U. acknowledges Russian Science Foundation (Project No 14-22-00281 ``Low dimensional quantum field theory in elementary particles and condensed matter physics''). I.T. acknowledges Skolkovo Foundation (Grant agreement for Russian educational and scientific organization no. 4dd.25.12.2014 to I.T.). A.N. acknowledges financial support from the EPSRC Established Career fellowship of A.V. Kavokin. The authors are grateful to M.V. Dubina for his attention to the work, to A.V. Syromyatnikov and A.G. Yashenkin for valuable discussions and to E.D. Eidelman for his support.

\end{acknowledgments}

\appendix

\section{ZA phonons self-energy part corrections}
\label{append}

From \eqref{pertZA} we see that in the second order in disorder strength we have different corrections. We treat this perturbation conventionally, considering only the terms with disorder constants corresponds to the same site. Thus, we have four terms, proportional to  $\langle \alpha^2_l \rangle$, $\langle \beta^2_{lj} \rangle$, $\langle \alpha_l \beta_{lj} \rangle$ and $\langle \beta_{lj} \beta_{lm} \rangle$ for $j \neq m$.

Correction proportional to $\langle \alpha^2_l \rangle$ is logarithmically divergent at very small momenta $k$ and contains the following integral
\begin{equation}
  I_1(k)=\frac{1}{2}\int \frac{d^2 q}{(2 \pi)^2}\frac{1}{(\omega^T_k)^2-(\omega^T_q)^2+i0}
\end{equation}

The second correction stems from nonzero average $\langle \beta^2_{lj} \rangle$. Corresponding equation is
\begin{equation}
  \sum_j \frac{32 \langle \beta^2_{lj} \rangle v_0 \sin^2{\frac{k_j}{2}}}{m^2 \omega^T_k} \int \frac{d^2 q}{(2 \pi)^2} \frac{\sin^2{\frac{q_j}{2}}}{(\omega^T_k)^2-(\omega^T_q)^2+i0},
\end{equation}
where summation is over three possible neighbors positions and $q_j$ and $k_j$ are corresponding momenta projections. We denote corresponding integral as
\begin{equation}
  I_{2l}(k)=\int \frac{d^2 q}{(2 \pi)^2} \frac{\sin^2{\frac{q_j}{2}}}{(\omega^T_k)^2-(\omega^T_q)^2+i0}.
\end{equation}
The third correction is due to correlations between  $\alpha_l$ and $\beta_{lj}$. Corresponding equation is
\begin{equation}
  \frac{2v_0 \langle \alpha_l \beta_{lj} \rangle}{m^2 \omega^T_k} \sum_j \int \frac{d^2 q}{(2 \pi)^2} \frac{f({\bf k}, {\bf q},j)}{(\omega^T_k)^2-(\omega^T_q)^2+i0},
  \label{corr21}
\end{equation}
where
\begin{equation}
  f({\bf k}, {\bf q},j)=1+\cos{(k_j-q_j)}-\cos{k_j}-\cos{q_j}.
\end{equation}
Expression \eqref{corr21} can be simplified using
\begin{multline}
  1+\cos{(k_j-q_j)}-\cos{k_j}-\cos{q_j}=\sin{k_j}\sin{q_j}\\+4\sin^2{\frac{k_j}{2}}\sin^2{\frac{q_j}{2}},
\end{multline}
the first term gives zero after integration over angle, so the correction obeys following form
\begin{equation}
  \frac{8v_0 \langle \alpha_l \beta_{lj} \rangle}{m^2 \omega^T_k} \sum_j \sin^2{\frac{k_j}{2}} \int \frac{d^2 q}{(2 \pi)^2} \frac{\sin^2{\frac{q_j}{2}}}{(\omega^T_k)^2-(\omega^T_q)^2+i0},
  \label{corr22}
\end{equation}
corresponding integral denotation is
\begin{equation}
  I_{3j}(k)= \int \frac{d^2 q}{(2 \pi)^2} \frac{\sin^2{\frac{q_j}{2}}}{(\omega^T_k)^2-(\omega^T_q)^2+i0}.
\end{equation}

The last correction arises from correlations between $\beta_{ij}$ and $\beta_{il}$, where $j$ and $l$ denote different neighboring sites. Corresponding equation is
\begin{equation}
    \sum_{j\neq m} \frac{32 \langle \beta_{lj} \beta_{lm} \rangle v_0 }{m^2 \omega^T_k} \int \frac{d^2 q}{(2 \pi)^2} \frac{g({\bf k}, {\bf q},j,m)}{(\omega^T_k)^2-(\omega^T_q)^2+i0},
\end{equation}
where
\begin{multline}
  g({\bf k}, {\bf q},j,m)=\cos{\frac{({\bf k}-{\bf q})({\bf e}_m-{\bf e}_j)}{2}} \\ \times\sin{\frac{k_j}{2}}\sin{\frac{k_m}{2}}\sin{\frac{q_j}{2}}\sin{\frac{q_m}{2}},
\end{multline}
so
\begin{equation}
  I_{4jm}(k)=\int \frac{d^2 q}{(2 \pi)^2} \frac{g({\bf k}, {\bf q},j,m)}{(\omega^T_k)^2-(\omega^T_q)^2+i0}.
\end{equation}

Real parts of this expressions are the principal values of corresponding integrals, and imaginary parts are
\begin{multline}
  \Im \left[\int \frac{d^2 q}{(2 \pi)^2} \frac{F({\bf k}, {\bf q})}{k^2-q^2+i0}\right]= \\ - \frac{\pi}{(2 \pi)^2} \int d\varphi \frac{ k F({\bf e}_k, {\bf e}_q,q=k)}{\frac{\partial (\omega^T_q)^2}{\partial q}(k)}
\end{multline}

The obtained with MD simulations values of energetic parameters are $\sqrt{<\alpha_l \beta_{lj}>}=\sqrt{-0.016}$\,eV/\AA$^2$ and $\sqrt{<\beta_{lj} \beta_{lm}>}=\sqrt{-0.002}$\,eV/\AA$^2$.

\section{Approximation of integral in the in-plane phonon self-energy part correction}
The integral in Eq. \eqref{sigma2LATAxiterm} required for calculating $\Sigma^{(2)}_{\xi}$ and $\tau_{\mathrm{substr}}$ for TA phonons can be approximated with sufficient accuracy as follows:
\begin{multline}
 \sum_j \sin^2{\frac{k_j}{2}} \\ \times \int \frac{d^2 q}{(2 \pi)^2} \frac{((\mathbf{p}_{k}-(\mathbf{p}_{k}\cdot\mathbf{e}_{j})\mathbf{e}_{j})\cdot(\mathbf{p}_{q}-(\mathbf{p}_{q}\cdot\mathbf{e}_{j})\mathbf{e}_{j}))^2\sin^2{\frac{q_j}{2}}}{k^2-q^2+i0} \\ \approx 0.2k^2-2.2k^3+2.33k^4-i(1.51k^4-1.11k^5).
\end{multline}
Its dependence on the direction of $\mathbf{k}$ with respect to graphene bonds is weak. For LA phonons the integral is four times larger.

\bibliography{graphene}

\end{document}